\documentclass[acmtog,noacm]{acmart}
\renewcommand\footnotetextcopyrightpermission[1]{} 
\setcopyright{none}
\usepackage{float}
\usepackage{subfigure}
\usepackage{multirow}
\usepackage{subcaption}
\usepackage{verbatim}
\usepackage{stfloats}
\usepackage{enumitem}
\AtBeginDocument{%
  \providecommand\BibTeX{{%
    \normalfont B\kern-0.5em{\scshape i\kern-0.25em b}\kern-0.8em\TeX}}}
\usepackage{colortbl}
\usepackage{makecell}
\newcommand{\pll}{\kern 0.56em/\kern -0.8em /\kern 0.56em}
\usepackage{bbding}
\usepackage{soul}

\newcommand{\spatialpnt}{\mathbf{p}}
\newcommand{\Othoframe}{\mathbf{V}}
\newcommand{\othoframe}{\mathbf{v}}
\newcommand{\symmetryfunction}{F_{\Othoframe}}

\newcommand{\shcoeff}{\mathbf{q}}
\newcommand{\direction}{\mathbf{d}}
\newcommand{\featuredirection}{\mathbf{d}_f}
\newcommand{\featureedges}{\mathcal{F}}
\newcommand{\diffofframes}{\mathcal{D}}

\newcommand{\tilderotate}[2]{\tilde{\mathbf{R}}_{{#1}\to {#2}}}
\newcommand{\tetcenters}{\mathcal{P}}
\newcommand{\boundarytetcenters}{\bar{\mathcal{P}}}
\newcommand{\dualedges}{\mathcal{E}}

\begin{document}
\title{NeurFrame: Learning Continuous Frame Fields for Structured Mesh Generation}

\author{Xiaoyang Yu}
\email{19020231153653@stu.xmu.edu.cn}
\author{Canjia Huang}
\email{huangcanjia0214@gmail.com}
\author{Zhonggui Chen}
\email{chenzhonggui@xmu.edu.cn}
\author{Juan Cao}
\authornote{Corresponding author: Juan Cao (juancao@xmu.edu.cn).}
\email{juancao@xmu.edu.cn}
\affiliation{
  \institution{Xiamen University}
  \country{China}
}

\begin{abstract}
Structured meshes, composed of quadrilateral elements in 2D and hexahedral elements in 3D, are widely used in industrial applications and engineering simulations due to their regularity and superior accuracy in finite element analysis. Generating high-quality structured meshes, however, remains challenging, especially for complex geometries and singularities. Field-guided approaches, which construct cross fields in 2D and frame fields in 3D to encode element orientation, are promising but are typically defined on discrete meshes, limiting continuity and computational efficiency. To address these challenges, we introduce \emph{NeurFrame}, a neural framework that represents frame fields continuously over the domain, supporting infinite-resolution evaluation. Trained in a self-supervised manner on discrete mesh samples, NeurFrame produces smooth, high-quality frame fields without relying on dense tetrahedral discretizations. The resulting fields simultaneously guide high-quality quadrilateral surface meshes and hexahedral volumetric meshes, with fewer and better-distributed singularities. By using a single network, NeurFrame also achieves lower computational cost compared to prior self-supervised neural methods that jointly optimize multiple fields.
\end{abstract}


\begin{teaserfigure}
    \centering
    \includegraphics[width=\textwidth]{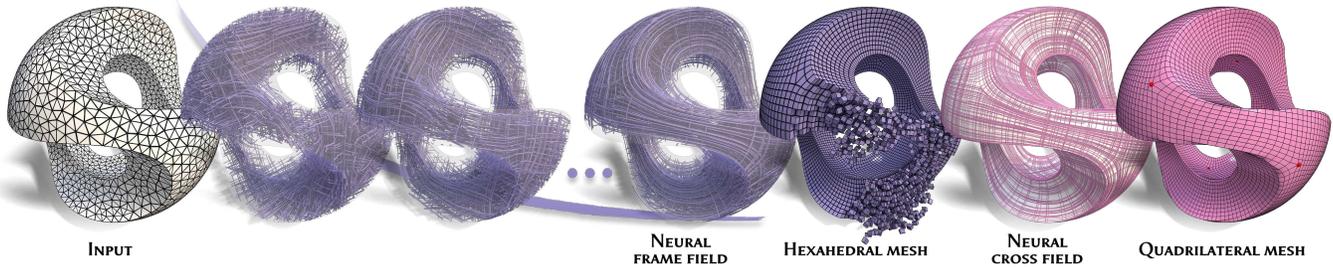}
    \caption{Continuous neural frame field generation for structured meshing. Starting from a randomly initialized volumetric frame field, NeurFrame progressively smooths it and aligns it with surface normals and specified features. The optimized field can drive high-quality hexahedral mesh generation and, via surface cross fields, produce high-quality quadrilateral meshes.}
    \label{fig:teaser}
\end{teaserfigure}

\maketitle
\thispagestyle{empty}

\section{INTRODUCTION}
\label{sec:introduction}

A structured mesh consists of regular elements: quadrilateral elements with four corner nodes in 2D and hexahedral elements with eight corner nodes in 3D \cite{Yeoh-2010}. Their regular cells and superior performance in finite element analysis make structured meshes widely adopted in industrial applications and engineering simulations, including polygonal modeling, texture mapping, computational fluid dynamics, tensor-product spline fitting, and isogeometric analysis \cite{Hughes-2005}.

However, generating structured meshes remains highly complex. In 2D, achieving the minimum number of singular points with optimal placement in quadrilateral mesh generation is still an open and active research problem \cite{Feng-2021}. In 3D, the challenges increase substantially, especially in constructing volumetric mappings that are locally injective, low-distortion, and compatible with integer constraints \cite{Pietroni-2022}. Despite decades of research and notable advances, algorithmically generating high-quality hexahedral meshes that conform to complex geometries remains largely unsolved \cite{Pietroni-2022}.

\begin{figure}[t]
    \centering
    \includegraphics[width=\columnwidth]{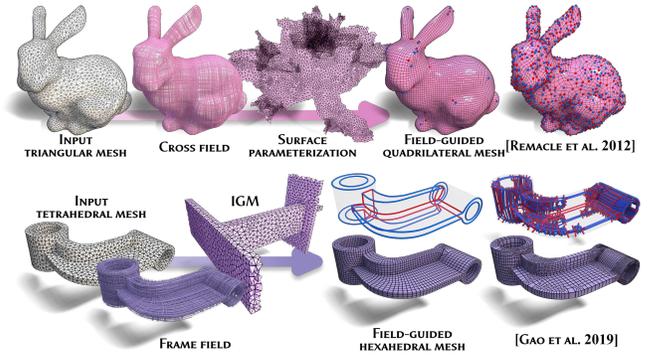}
    \caption{\textcolor{black}{The field-guided structured mesh generation pipeline begins with an input triangular or tetrahedral mesh. A cross or frame field is optimized on the surface or volume to guide parameterization, aligning the mesh elements with the field directions. The structured mesh is then extracted from this parameterization. Compared to other quadrilateral or hexahedral mesh generation methods, such as the triangle-merge~\cite{Remacle-2012} and octree-based~\cite{Gao-2019} approaches, field-guided methods yield higher-quality meshes with fewer singularities (red and blue) and a more balanced singularity distribution.}}
    \label{fig:field-guided-meshing-pipeline}
\end{figure}

Field-guided methods are a promising class of techniques for structured mesh generation. They first construct a guiding field called a cross field in 2D or a frame field (also referred to as an \emph{octahedral field}) in 3D. This field encodes the desired local orientation and structural properties of mesh elements. In 2D \textcolor{black}{(top of Fig.~\ref{fig:field-guided-meshing-pipeline})}, the cross field drives surface parameterization \cite{Kälberer-2007, Bommes-2009, Myles-2014}, from which a regular quadrilateral tessellation in the parametric domain is pulled back to the surface \cite{Ebke-2013}. This produces a structured quadrilateral mesh. In 3D \textcolor{black}{(bottom of Fig.~\ref{fig:field-guided-meshing-pipeline})}, the frame field guides the construction of an integer-grid map (IGM) \cite{Nieser-2011, Brückler-2022a}. The IGM is an atlas of charts whose maps and transition functions satisfy boundary alignment, local injectivity, and conformity. A regular cubic tessellation is then pulled back to the volume, resulting in a high-quality hexahedral mesh \cite{Lyon-2016, Kohler-2025}.

Compared with other quadrilateral or hexahedral mesh generation techniques, field-guided approaches can represent arbitrary singularity structures; see Fig.~\ref{fig:field-guided-meshing-pipeline}. This flexibility enables the construction of high-quality meshes, especially when precise alignment with complex boundary features and internal structures is required~\cite{Pietroni-2022}. Despite their effectiveness, constructing high-quality cross and frame fields remains challenging. Most existing methods operate on discrete surface or volumetric meshes and solve for the fields via optimization on these discretizations. As a result, the resulting fields are defined only at discrete locations. In 3D, tetrahedral discretizations further introduce a trade-off between computational cost and field quality. Fine tetrahedral meshes are expensive to generate and process, whereas coarse meshes suffer from discretization artifacts that lead to noisy and irregular field topology \cite{Solomon-2017}. A continuous frame field representation can potentially reduce computational cost while maintaining high-quality fields by decoupling optimization from dense tetrahedral meshes.

Based on the above observations, we propose a neural approach for computing frame fields. Our method represents the frame field as a continuous neural function trained from discrete samples and uses it to guide structured mesh generation. As shown in Fig.~\ref{fig:teaser}, by exploiting the cubic symmetry and boundary alignment properties of frame fields, our approach supports both high-quality hexahedral and quadrilateral mesh generation. Compared to traditional field-based methods, it avoids reliance on fine tetrahedral discretizations and directly produces a continuous frame field with fewer and better-distributed singularities. Moreover, it is more efficient than existing neural field-learning approaches. We call this method \emph{NeurFrame} and summarize our contributions below.

\begin{itemize}[leftmargin=*]
\item We propose \emph{NeurFrame}, a self-supervised neural approach that provides a continuous representation of frame fields, allowing optimization on coarse volumetric discretizations while supporting queries throughout the entire model.
\item Leveraging the strong approximation capability of neural networks, NeurFrame produces frame fields that simultaneously guide high-quality quadrilateral surface meshing and hexahedral volumetric meshing, with fewer and better-distributed singularities, improving mesh regularity and structural fidelity.
\item NeurFrame employs a single network, thereby achieving lower computational cost than prior self-supervised neural methods that jointly optimize two neural fields, including cross-field method \cite{Dong-2025b} and frame-field method~\cite{Zheng-2025}.
\end{itemize}

\section{RELATED WORK}
\label{sec:related-work}
\thispagestyle{fancy}
\fancyfoot{}

\textcolor{black}{
Structured mesh generation has been extensively studied for several decades \cite{Bommes-2013, Pietroni-2022}. Among the many existing methods, field-guided approaches offer distinctive advantages, as discussed in Sec.~\ref{sec:introduction}. Rather than reviewing all structured meshing techniques, we focus on field-based methods that construct cross fields in 2D and frame fields in 3D to guide quadrilateral and hexahedral mesh generation.
}

\subsection{Cross Field Generation for Quadrilateral Meshing}
Cross fields guide quadrilateral mesh generation. At each point, a cross contains two orthogonal directions and exhibits 4-rotational symmetry (4-RoSy) \cite{Ray-2008, Lai-2010}. They are generally optimized to maximize smoothness while aligning with specified features or principal directions. For example, \citet{Knoppel-2013} minimize a quadratic smoothness energy to construct cross fields on surfaces. \citet{Diamanti-2014} introduce N-PolyVector fields, representing N-RoSy fields as roots of complex polynomials and solving sparse linear systems, later extended to curl-free fields in \cite{Diamanti-2015}. \citet{Jakob-2015} generate quad-dominant meshes by locally optimizing cross fields, often producing many singularities. \citet{Huang-2018} improve field-guided quadrilateral meshing by combining linear and quadratic constraints to reduce singularities. \citet{Viertel-2019} compute boundary-aligned cross fields by minimizing a Ginzburg--Landau energy using MBO threshold dynamics. \citet{Zhang-2020} propose a frame-field-based method to design smooth cross fields that automatically align with sharp features. \citet{Pietroni-2021} combine cross field design with constrained patch layouts to construct feature-aligned quadrilateral meshes, sometimes causing severe distortion in irregular patches.

Recently, several neural methods have been proposed for cross field generation. Data-driven approaches learn cross fields from existing meshes or from precomputed cross-field data: \citet{Dielen-2021} predict cross fields by training on a collection of quad meshes, while CrossGen \cite{Dong-2025a} supports both feed-forward prediction and latent generative modeling. In contrast, the self-supervised method NeurCross \cite{Dong-2025b} does not rely on ground-truth data. It jointly optimizes a neural signed distance field (SDF) and a neural cross field to align the field with principal directions. However, the SDF is reliable only near the surface, limiting applicability for volumetric frame field prediction, and the complex architecture and joint optimization incur high computational cost.

\subsection{Frame Field Generation for Hexahedral Meshing}

Not all frame fields are suitable for guiding IGMs to produce high-quality hexahedral meshes. A key requirement is smoothness in the volume interior, characterized by small spatial gradients, which ensures that mesh edge orientations vary smoothly~\cite{Huang-2011}. Assessing smoothness between frames, however, is challenging. Unlike 2D 4-symmetry direction fields, 3D frames involve 24 possible transformations within a noncommutative rotation group ($SO(3)$), complicating the direct extension of 2D representation and optimization methods.

\citet{Huang-2011} first proposed a spherical harmonics (SH) representation, encoding a frame as a function on the unit sphere and quantifying smoothness as an integral over the sphere. \citet{Ray-2016} improved initialization by enforcing boundary conditions, while \citet{Kowalski-2016} prioritized stable regions over boundaries. \citet{Corman-2019} used Cartan’s method of moving frames with the Darboux derivative as the main representation. \citet{Zhang-2020} introduced smoothness energies for SH-based frame fields. \citet{Palmer-2020} proposed odeco fields, allowing independent scaling along three orthogonal directions to better capture volumetric singularities, and \citet{Palmer-2021} discretized anisotropic fourth-order differential operators parameterized by symmetric frame fields. \citet{Fang-2023} introduced a metric-aware smoothness measure, enforcing local integrability and generating cubic-symmetry frame fields guided by Riemannian metrics and alignment constraints. To address non-meshability caused by incompatibilities between singularities and hexahedral topology, \citet{Liu-2018} derived necessary local and global conditions for frame-field hex-meshability, and \citet{Liu-2023} extended this by modifying non-meshable fields to be locally meshable.

All prior methods define cross fields or frame fields on discrete meshes, with values specified at mesh faces or vertices, or at tetrahedral centroids or vertices rather than as continuous functions over the domain. Consequently, dense discretizations are often required to obtain high-quality fields, resulting in high computational cost. \citet{Solomon-2017} introduced an infinite-resolution volumetric frame field via the boundary element method, but relied on a relax-and-project pipeline rather than direct optimization, limiting feature alignment and computational efficiency. More recently, \citet{Zheng-2025} proposed a neural octahedral field with a continuous representation for surface reconstruction, though its use for volumetric parameterization or hexahedral meshing remains unexplored. In contrast, we propose NeurFrame, a self-supervised neural method that learns a continuous volumetric frame field. It enables optimization on coarse discretizations while supporting dense queries throughout the domain, and it simultaneously guides high-quality quadrilateral surface meshing and volumetric hexahedral meshing with fewer singularities and lower computational cost.

\begin{figure*}[t]
\centering
\includegraphics[width=\textwidth]{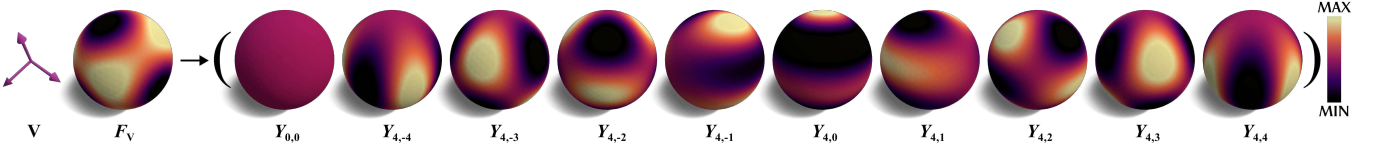}
\caption{Spherical harmonic (SH) representation of a frame. The leftmost image shows a frame $\Othoframe$ together with its associated spherical function $\symmetryfunction$ on the unit sphere, where the function minima indicate the frame directions. The remaining plots show the projection of $\symmetryfunction$ onto the SH basis functions $Y_{0,0}$ in band 0 and $Y_{4,i}$ in band 4, with $i=-4,\dots,4$. The color scale encodes function values, from minimum (dark) to maximum (bright).}
\label{fig:SH-representation-of-frame}
\end{figure*}

\section{PRELIMINARIES}
\thispagestyle{fancy}
\fancyfoot{}
\subsection{SH Representation of Frame Field}
For a spatial point $\spatialpnt\in\mathbb{R}^3$ in the domain, the associated frame can be represented by a matrix $\Othoframe=[\othoframe_0, \othoframe_1, \othoframe_2] \in \mathbb{R}^{3\times 3}$, where the mutually orthogonal unit vectors $\othoframe_0, \othoframe_1, \othoframe_2 \in \mathbb{R}^3$ form the frame axes, with $\othoframe_2 = \othoframe_0\times \othoframe_1$. \citet{Huang-2011} encode $\Othoframe$ as a function defined on the unit sphere $\mathcal{S}^2=\{\mathbf{s}\in\mathbb{R}^3 \mid \Vert\mathbf{s}\Vert_2=1\}$:
\begin{equation}
    \symmetryfunction(\mathbf{s})=
    (\othoframe_0^{\top}\mathbf{s})^2(\othoframe_1^{\top}\mathbf{s})^2 + 
    (\othoframe_1^{\top}\mathbf{s})^2(\othoframe_2^{\top}\mathbf{s})^2 + 
    (\othoframe_2^{\top}\mathbf{s})^2(\othoframe_0^{\top}\mathbf{s})^2,
    \mathbf{s}\in\mathcal{S}^2. 
    \notag
     \label{eq:original-formulation}
\end{equation}
This function is invariant under permutations and sign changes of the frame axes; for example, $F_{[\othoframe_0, \othoframe_1, \othoframe_2]} = F_{[\othoframe_0, \othoframe_2, -\othoframe_1]}$. The frame axes correspond to the directions of the minima of the spherical function $\symmetryfunction$~\cite{Huang-2011}; see the leftmost example in Fig.~\ref{fig:SH-representation-of-frame}. Over the unit sphere $\mathcal{S}^2$, \citet{Huang-2011} employ an orthogonal basis for spherical functions using Laplace–Beltrami eigenfunctions. For band $l$, the eigenspace is spanned by the SH $Y_{l,m}$, with $m=-l,\dots,l$. The function $\symmetryfunction$ can be projected losslessly onto the SH bases in bands 0 and 4~\cite{Huang-2011}, as illustrated in Fig.~\ref{fig:SH-representation-of-frame}.
\begin{equation}
    \symmetryfunction(\mathbf{s})= c_0Y_{0,0}+c_1\sum_{i=-4}^{4}q_i Y_{4,i}(\mathbf{s}), \mathbf{s}\in \mathcal{S}^2, \label{equ:lossess-projection}
\end{equation}
where $c_0$ and $c_1$ are constants determined by the projection, $\shcoeff=(q_{-4},\cdots,q_{4}) \in\mathbb{R}^9$ is the coefficient vector for SH bases in band 4, $Y_{0,0}$ denotes the real SH basis in band 0, and $Y_{4,-4}(\mathbf{s}), \ldots, Y_{4,4}(\mathbf{s})$ are the real SH bases in band 4 evaluated at $\mathbf{s}$. Consequently, a frame $\Othoframe$ can be indirectly represented by the coefficient vector $\shcoeff$. Since the SH basis functions form an othogonal basis on $\mathcal{S}^2$, the difference between two frames $\Othoframe^a$ and $\Othoframe^b$ can be expressed as $\diffofframes(\Othoframe^a, \Othoframe^b) = \Vert\shcoeff^a-\shcoeff^b\Vert_2^2$ \cite{Ray-2016, Zheng-2025}. Thus, comparing $\Othoframe^a$ and $\Othoframe^b$ reduces to comparing their corresponding SH coefficient vectors $\shcoeff^a$ and $\shcoeff^b$.

\subsection{Alignment of Frame}
During frame field optimization, boundary conditions often require one axis of the frame $\Othoframe$ to align with a prescribed direction $\direction \in \mathbb{R}^3$. For frames represented explicitly (e.g., $\Othoframe=[\othoframe_0, \othoframe_1, \othoframe_2]$), such alignment can be enforced by directly applying a rotation. However, for frames encoded by SH coefficients as in Eq.~(\ref{equ:lossess-projection}), this operation cannot be performed directly, and an indirect formulation is required to measure and enforce alignment.

Given an SH coefficient vector $\shcoeff$ and a unit direction $\direction$, \citet{Ray-2016} derive a corresponding $9 \times 9$ rotation matrix $\tilderotate{\direction}{\mathbf{z}}$ acting on the SH coefficient space. Applying $\tilderotate{\direction}{\mathbf{z}}$ to $\shcoeff$ yields a rotated coefficient vector $\tilderotate{\direction}{\mathbf{z}}\shcoeff$, which can then be compared to the family of SH representations corresponding to frames whose axes are aligned with the $z$-axis:
\begin{equation}
    \shcoeff_z=\left[
    \sqrt{\frac{5}{12}}\cos4\theta,0,0,0,\sqrt{\frac{7}{12}}, 0,0,0,\sqrt{\frac{5}{12}}\sin4\theta \notag
    \right],
\end{equation}
where $\theta$ parameterizes the in-plane rotation. As shown by \citet{Ray-2016}, enforcing
\begin{equation}
    \mathbf{e}_0^{\top}\tilderotate{\direction}{\mathbf{z}}\shcoeff=\sqrt{\frac{7}{12}} \label{sh-align}
\end{equation}
is sufficient to guarantee that the frame encoded by $\shcoeff$ has one axis aligned with the target direction $\direction$. Here, $\mathbf{e}_0 = (0,0,0,0,1,0,0,0,0)^{\top}$ denotes the basis vector corresponding to the coefficient of $Y_{4,0}$.

\section{NEURAL REPRESENTATION AND OPTIMIZATION OF FRAME FIELDS}
\label{sec:method}
Recent advances in deep learning have demonstrated the effectiveness of neural networks in modeling complex continuous functions. In particular, multilayer perceptrons (MLPs) with learnable parameters can approximate highly nonlinear mappings from spatial coordinates to function values, a capability supported by the universal approximation theorem \cite{Kim-2003, Lecun-2015}. Neural representations have been successfully applied to a variety of continuous fields \cite{Xie-2022}, including neural radiance fields \cite{Mildenhall-2021} and SDFs \cite{Park-2019, Ben-2022}. Inspired by the success of recent neural approaches, we propose using an MLP to represent a continuous frame field, which alleviates the performance bottlenecks of traditional methods based on discrete frame fields, especially when high-precision evaluations are needed. To the best of our knowledge, this is the first work to employ an MLP to represent a spatially continuous frame field for structured mesh generation. In this section, we describe both the neural representation and the loss functions used to optimize the network for producing high-quality frame fields.
\subsection{Neural Representation of Frame Field}

We represent the spatially varying frame field using an MLP tailored for volumetric parameterization and hexahedral mesh generation. The two frame axes orthogonal to the surface normal are further leveraged to guide quadrilateral meshing. Following \cite{Zheng-2025}, the frame field is parameterized using SH coefficients, which provide a compact, rotation-aware representation. Specifically, the MLP maps a 3D spatial position $\spatialpnt$ to a 9D SH coefficient vector $\shcoeff$:
\begin{equation}
    f(\spatialpnt;\Theta)=\shcoeff,\quad\spatialpnt\in\mathbb{R}^3, \shcoeff\in\mathbb{R}^9, 
    \label{eq:MLPmaps} \notag
\end{equation}
where $\Theta$ denotes the learnable network parameters. To restrict the solution space and enforce a necessary condition for producing a valid frame field \cite{Zheng-2025}, we normalize the network output such that $\|\shcoeff\|_2 = 1$.

\subsection{Neural Frame Field Optimization}
\label{subsec:optimization}
\subsubsection{Smoothness}

A fundamental requirement for frame fields in IGM-based hexahedral mesh generation is interior smoothness, which promotes gradual variations in edge orientations in the resulting mesh. To quantify smoothness, we adopt a discretization of the Dirichlet energy similar to prior work~\cite{Liu-2018}, but introduce
\begin{minipage}[b]{\dimexpr\columnwidth-0.2\columnwidth-0.05\columnwidth}
edge-length-based weights to better account for the spatial distribution of tetrahedra. Specifically, let $\tetcenters$ denote the centroids of all tetrahedra and let $\dualedges$ denote the set of dual edge connecting centroids between adjacent tetrahedra. The smoothness loss is then defined as:
\end{minipage}
\hfill
\begin{minipage}[b]{0.2\columnwidth}
    \centering
    \includegraphics[width=\linewidth]{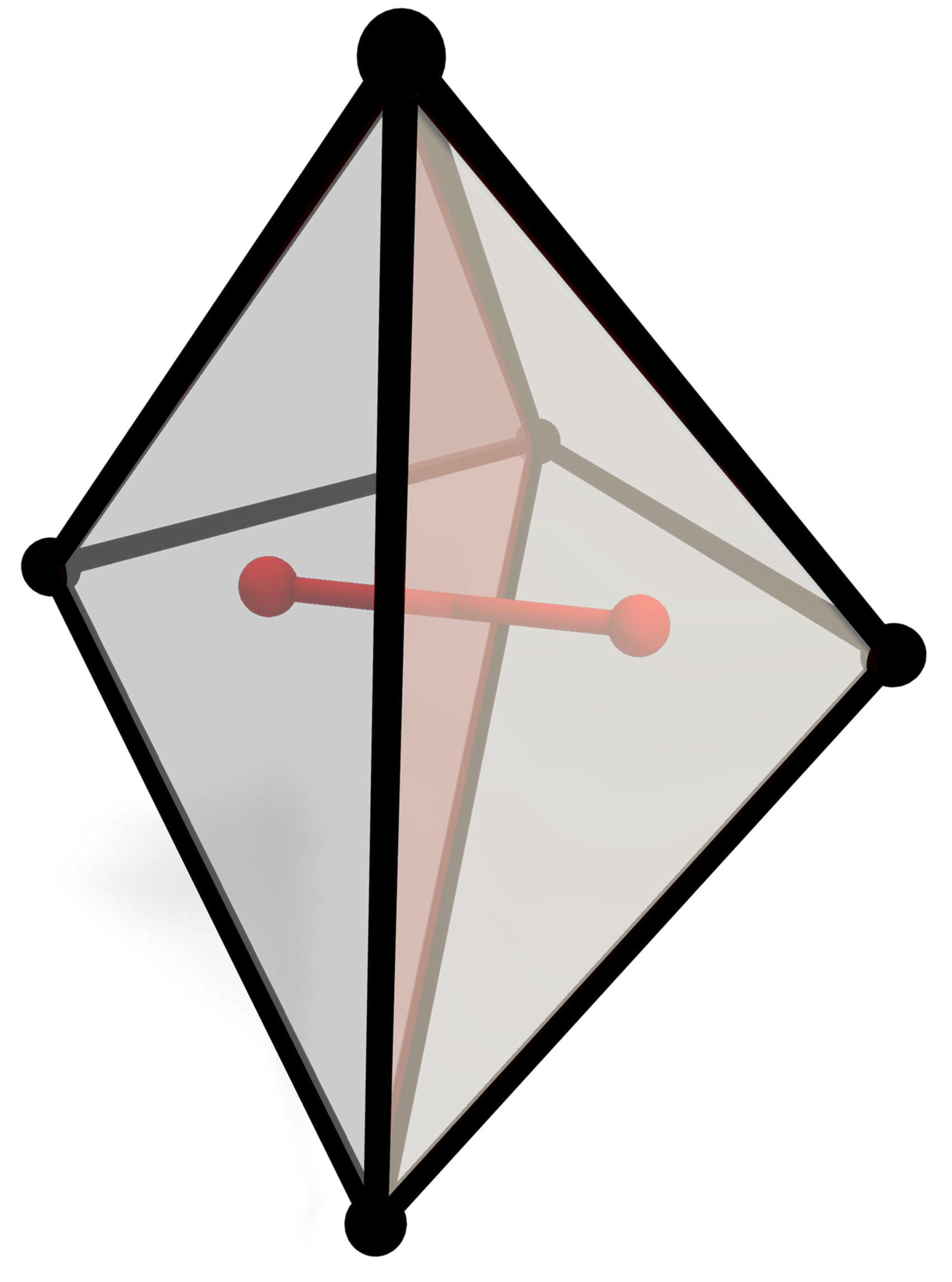}
\end{minipage}
\begin{equation}
    \mathcal{L}_{\mathrm{S}}=\frac{1}{|\dualedges|}\sum_{\spatialpnt\hat{\spatialpnt}\in\dualedges} \omega_{\spatialpnt\hat{\spatialpnt}}\Vert f(\spatialpnt;\Theta) - f(\hat{\spatialpnt};\Theta)\Vert_2^2 , \label{loss:smooth}
\end{equation}
where $\omega_{\spatialpnt\hat{\spatialpnt}}=\frac{1}{\|\spatialpnt\hat{\spatialpnt}\|+\tau}$ is a weight associated with the length of the dual edge connecting $\spatialpnt$ and $\hat{\spatialpnt}$, and to prevent degeneracy, we set $\tau=10^{-2}$.

\subsubsection{Boundary Alignment}

Another essential requirement for frame fields in hexahedral mesh generation is boundary alignment. Specifically, on the boundary of the volume, the frame field is constrained such that one of its axes aligns with the surface normal. Similar to the smoothness term, this constraint is enforced at the centroids of tetrahedra. For a boundary tetrahedron, the target normal is taken as the normal of its associated boundary face. To avoid conflicting alignment constraints where a single centroid would need to align with multiple boundary normals, we subdivide tetrahedra that contain more than one boundary face. This ensures that each tetrahedron corresponds to at most one boundary face.
We denote the set of centroids of boundary tetrahedra by $\boundarytetcenters$, where each point $\spatialpnt \in \boundarytetcenters$ is associated with a boundary normal $\mathbf{n}$. Following the SH-based alignment measure of \citet{Ray-2016} in Eq.~\eqref{sh-align}, we define the boundary alignment loss as:
\begin{equation}
    \mathcal{L}_{\mathrm{B}}=\frac{1}{|\boundarytetcenters|}\sum_{\spatialpnt\in\boundarytetcenters} \left(
    \sqrt{\frac{7}{12}} - \mathbf{e}_0^{\top}\tilderotate{\mathbf{n}}{\mathbf{z}}f(\spatialpnt;\Theta)
    \right)^2. 
    \label{loss:align-normal}
\end{equation}

\subsubsection{Feature Alignment}
\thispagestyle{fancy}
\fancyfoot{}

Our neural network optimization is defined on tetrahedral centroids and dual edges. However, concave feature edges do not induce dual edges between their adjacent tetrahedra. As a result, smoothness alone is insufficient to propagate alignment constraints across such regions. We therefore explicitly enforce the frame field to align with the model’s feature edges, whose set is denoted by $\featureedges$. Due to the cubic symmetry of the spherical harmonic representation, feature alignment is enforced by constraining one axis of the local frame to align with the feature direction, following the same principle as the boundary normal alignment in Eq.~\eqref{loss:align-normal}. To respect the locality of feature constraints, we further introduce a distance-weighted formulation that enforces alignment only in the vicinity of feature edges and avoids overconstraining the field elsewhere.

Let $\featuredirection \in \mathbb{R}^3$ denote the direction of the feature edge closest to the tetrahedral centroid $\spatialpnt$. The feature alignment loss is defined as:
\begin{equation}
    \mathcal{L}_{\mathrm{F}}=\frac{1}{|\tetcenters|}\sum_{\spatialpnt\in\tetcenters} \sigma_{\spatialpnt}\left(
    \sqrt{\frac{7}{12}} - \mathbf{e}_0^{\top}\tilderotate{\featuredirection}{\mathbf{z}}f(\spatialpnt;\Theta)
    \right)^2, \label{loss:align-feature}
\end{equation}
where $\sigma_{\spatialpnt}=\exp(-\sigma \cdot d(\spatialpnt, \featureedges))$. Here, $\sigma$ is a user-defined parameter that controls the spatial extent and strength of the feature alignment constraint. Larger values of $\sigma$ enforce a more localized influence around feature edges, while smaller values allow the constraint to propagate farther into the interior. In all our experiments, we set $\sigma=10$, which we found to provide a good balance between local feature preservation and global field smoothness. The function $d(\spatialpnt, \featureedges)$ denotes the Euclidean distance from $\spatialpnt$ to its nearest feature edge in $\featureedges$. This spatially decaying weighting encourages frames near feature edges to align with the corresponding feature direction, while gradually relaxing the constraint for frames farther away.

\subsubsection{Total Loss}
\textcolor{black}{
By combining the loss terms defined in Eqs.~\eqref{loss:smooth}-\eqref{loss:align-feature}, we formulate the overall objective as:
\begin{equation}
    \mathcal{L} = \lambda_{\mathrm{S}}\mathcal{L}_{\mathrm{S}}  + \lambda_{\mathrm{B}}\mathcal{L}_{\mathrm{B}} + \lambda_{\mathrm{F}}\mathcal{L}_{\mathrm{F}}, \label{loss:total}
\end{equation}
where $\lambda_{\mathrm{S}}$, $\lambda_{\mathrm{B}}$ and $\lambda_{\mathrm{F}}$ are weighting coefficients that balance the contributions of the smoothness, boundary alignment, and feature alignment terms, respectively. 
}

\section{EXPERIMENTS}
\label{sec:experiments}
\thispagestyle{fancy}
\fancyfoot{}

NeurFrame computes a continuous neural frame field throughout the volume. Unlike cross fields optimized only on surfaces, a volumetric frame field naturally aligns with sharp geometric features \cite{Zhang-2020}. This property allows NeurFrame to be evaluated on both hexahedral and quadrilateral mesh generation.

\subsection{Implementation Details}

We adopt a SIREN~\cite{Sitzmann-2020} architecture to represent the neural frame field, following its widespread use in continuous neural field modeling~\cite{Ben-2022, Zheng-2025}. The network consists of four hidden layers, each with 256 units. In all experiments, we set the weighting coefficients in Eq.~\eqref{loss:total} to $\lambda_{\mathrm{S}} = 1$, $\lambda_{\mathrm{B}} = 20$, and $\lambda_{\mathrm{F}} = 1$ by default. The model is optimized using the Adam optimizer~\cite{Kingma-2014} with a learning rate of $5\times10^{-5}$ for 10k iterations. NeurFrame takes a volumetric tetrahedral mesh as input. When only a surface mesh is available, we employ fTetWild~\cite{Hu-2020} to tetrahedralize the enclosed volume. The target edge length of the tetrahedra is set to $0.025$ times the bounding box diagonal of the input surface mesh.

All inputs are normalized to the range $[-1,1]^3$, and the network outputs are normalized to unit vectors in $\mathbb{R}^9$. After optimization, the trained network represents a continuous frame field and allows querying SH coefficients $\shcoeff$ at arbitrary spatial locations. These coefficients are then projected onto the nearest valid rotation matrix using the linearized projection method proposed by \citet{Ray-2016}, yielding the explicit frame $\Othoframe$. Our experimental datasets are drawn from Thingi10K~\cite{Zhou-2016}, HexMe~\cite{Beaufort-2022}, and additional models collected from online repositories. All experiments are conducted on a single NVIDIA GeForce RTX 3090 GPU with 24~GB of VRAM, running Ubuntu 20.04.6 LTS ($\mathrm{x}86\_64$). We develop algorithms for better visualizing our continuous frame field and its singularity graph, details in the Supplementary Material.

\begin{figure}[t]
    \centering
    \includegraphics[width=\columnwidth]{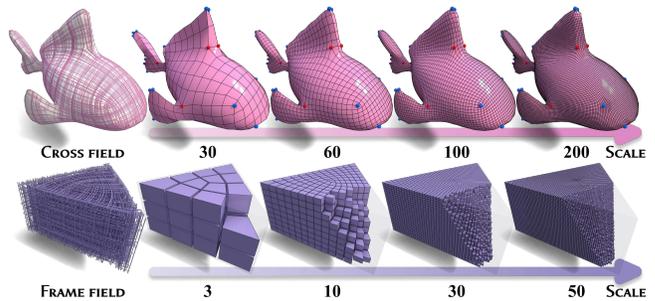}
    \caption{\textcolor{black}{Scaling the input surface or volumetric mesh changes the number of quadrilateral or hexahedral elements while preserving the singularity structure defined by the cross or frame field.
}}
    \label{fig:different-scale-meshing}
\end{figure}

\subsubsection{Pipeline for Quadrilateral Mesh Generation}

As our focus is on learning a high-quality continuous frame field rather than developing a complete structured meshing pipeline, we adopt standard techniques for quadrilateral mesh extraction. After optimization, we query the frame field at the centroid of each triangle of the input surface mesh. By discarding the axis most aligned with the triangle normal, we obtain a discrete cross field defined over the surface. We then compute a surface parameterization guided by this cross field using mixed-integer quadrangulation (MIQ)~\cite{Bommes-2009}. The final quadrilateral mesh is extracted from the parameterization using libQEx~\cite{Ebke-2013}. The resolution of the resulting mesh is controlled by adjusting the global scaling factor during parameterization, as shown in the top of Fig.~\ref{fig:different-scale-meshing} (default: 100). 

\subsubsection{Pipeline for Hexahedral Mesh Generation}

Given the optimized continuous neural frame field, we first discretize it by querying the frame at the centroid of each tetrahedron in the input volumetric mesh. Based on this discretized field, we compute a frame-aligned IGM using CubeCover~\cite{Nieser-2011}. The final hexahedral mesh is then extracted from the IGM using libHexEx~\cite{Lyon-2016}. The resolution of the resulting hexahedral mesh can be controlled by scaling the model during the parameterization stage. This allows us to generate meshes with varying cell counts, as illustrated in the bottom of Fig.~\ref{fig:different-scale-meshing} (default: 30). 

\subsection{Comparisons}

We compare our NeurFrame with a broad set of existing cross field and frame field generation techniques, as well as field-guided quadrilateral and hexahedral meshing methods. The most closely related competing method is Neural Octahedral Field (NeurOcta)~\cite{Zheng-2025}, which also produces a neural frame field. Since NeurOcta operates on points sampled from the surface, we uniformly sample 50k points from the input tetrahedral mesh boundary by~\cite{Corsini-2012} and provide them as input.

\subsubsection{Field-guided Quadrilateral Mesh Generation}

We benchmark NeurFrame against state-of-the-art cross field–guided quadrilateral meshing methods, including InstantMeshes~\cite{Jakob-2015}, QuadriFlow~\cite{Huang-2018}, and QuadWild~\cite{Pietroni-2021}. We also compare against surface parameterization–based approaches for quadrilateral meshing, namely MIQ and rectangular surface parameterization (RSP)~\cite{Corman-2025}. In addition, we include several cross-field generation methods, including power fields~\cite{Knoppel-2013}, polyvector fields~\cite{Diamanti-2014}, integrable fields~\cite{Diamanti-2015}, and Ginzburg–Landau fields~\cite{Viertel-2019}, all implemented using the Directional library~\cite{Amir-2025}. Neural-based field generation methods, NeurCross~\cite{Dong-2025b} and NeurOcta, are also included in the comparison. Similar to our approach, NeurOcta produces a volumetric frame field. For surface-based comparisons, we extract a cross field by querying the frame at the centroid of each surface triangle. For methods that output only a surface cross field or a surface parameterization, we apply the same quadrilateral extraction pipeline and parameter settings as in NeurFrame to ensure consistency.

\begin{figure}[t]
    \centering
    \includegraphics[width=\columnwidth]{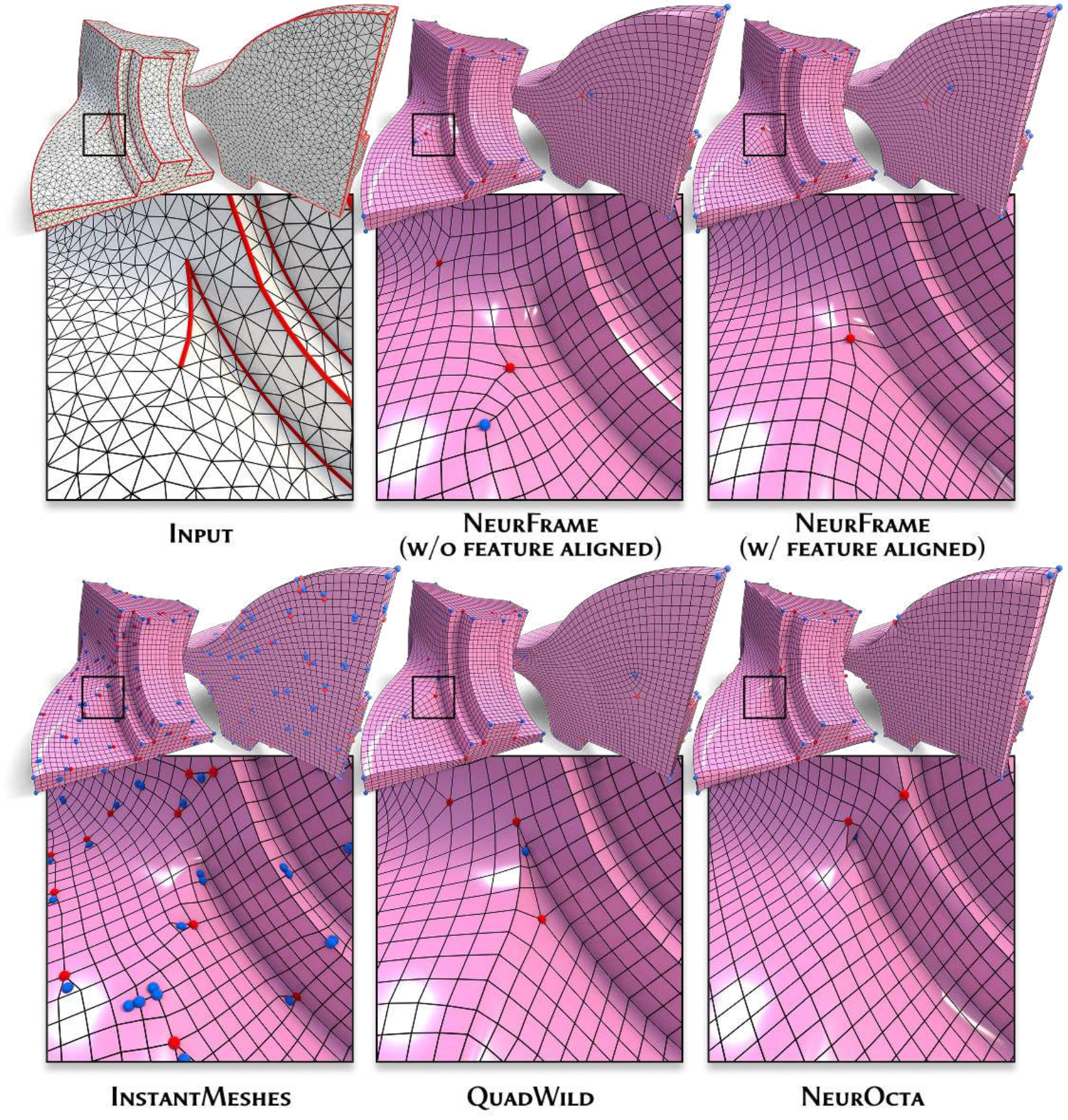}
    \caption{Comparison of quadrilateral meshes generated by feature-aligned methods. Feature edges are identified as mesh edges whose adjacent face normals differ by more than $\pi/4$ and are marked in red. Singular points, i.e., vertices with valence not equal to four, are shown in red (higher valence) and blue (lower valence).}
    \label{fig:comparison-quad-meshing-features}
    \vspace{-0.3cm} 
\end{figure}

To ensure fairness when comparing methods whose performance depends on the output mesh resolution, we normalize the target face count across methods. Specifically, we use the output of QuadWild as a reference, which automatically determines its target edge length based on the input surface area, set to $10^{-4}$ times the input surface area. The qualitative comparison is shown in Fig.~\ref{fig:comparison-quad-meshing}. Although NeurFrame does not explicitly optimize for alignment with principal curvature directions, it benefits from the cubic symmetry of the frame field representation and the enforced normal alignment constraint in Eq.~\eqref{loss:align-normal}. As a result, NeurFrame effectively captures local geometric features and produces quadrilateral meshes with improved directional coherence. In particular, the generated elements exhibit better alignment with dominant surface directions and contain fewer, more evenly distributed singularities compared to competing approaches.

\begin{figure}[t]
    \centering
    \includegraphics[width=\columnwidth]{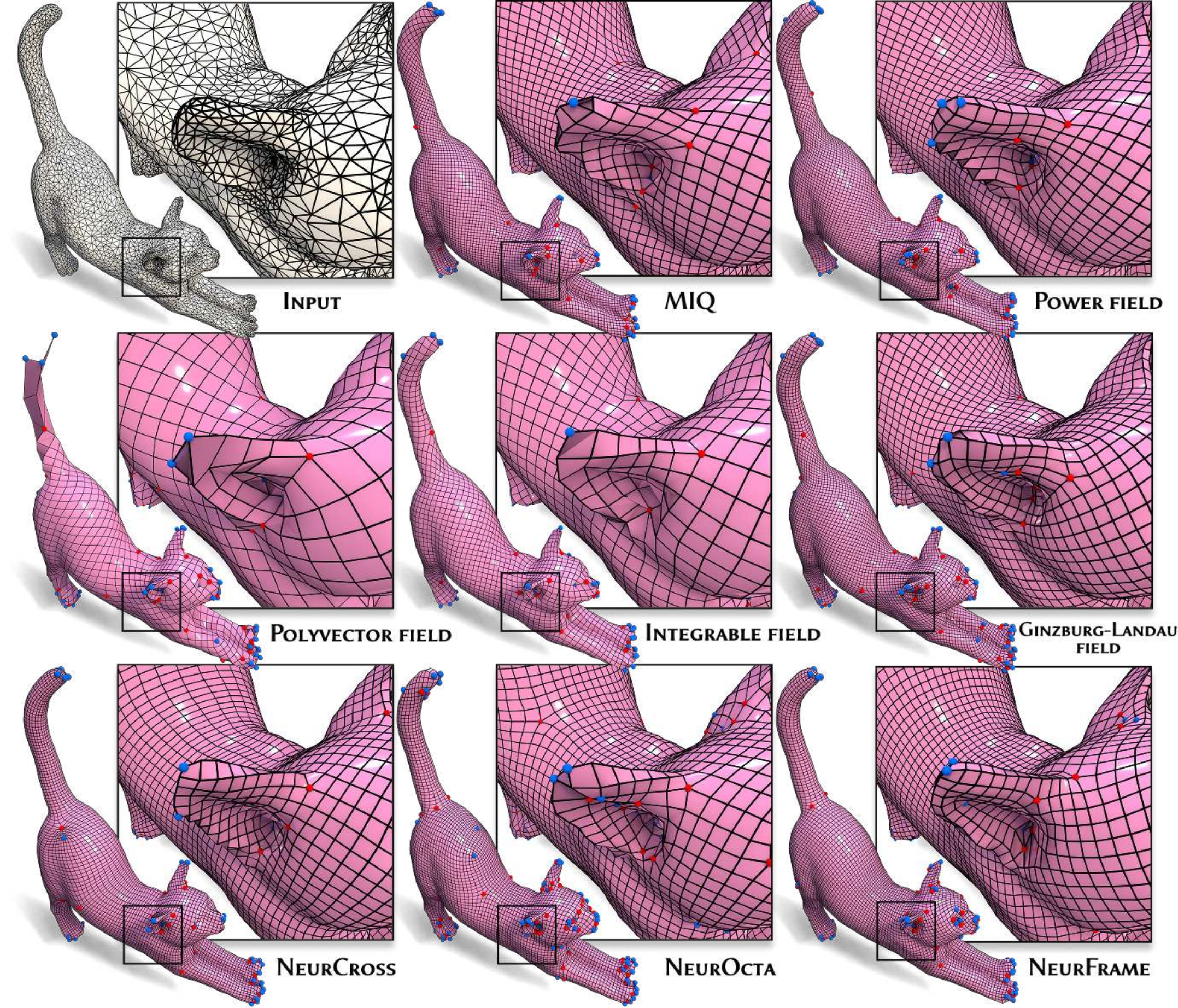}
    \caption{Comparison of quadrilateral meshes guided by cross fields generated using different methods. In challenging saddle-shaped regions, such as the cat’s back, only NeurCross achieves alignment comparable to NeurFrame. For fine-scale features, including the cat’s ears, only the polyvector field and Ginzburg-Landau field produce cross fields with alignment quality similar to ours.}
    \label{fig:comparison-quad-meshing-cross}
    \vspace{-0.3cm} 
\end{figure}
 
We further compare NeurFrame with feature-preserving methods in Fig.~\ref{fig:comparison-quad-meshing-features}. Even without explicitly enforcing the feature alignment term in Eq.~\eqref{loss:align-feature}, NeurFrame naturally aligns with sharp features, since the frame field simultaneously respects the surface normals on both sides of a crease. For shallow or subtle features, which may be difficult to capture implicitly (see the close-up in Fig.~\ref{fig:comparison-quad-meshing-features}), we explicitly detect these features and impose corresponding alignment constraints. 
InstantMeshes relies solely on extrinsic smoothness, aligning only relatively sharp features. QuadWild pre-segments the surface along sharp edges, often causing severe distortion within irregular patches. NeurOcta achieves limited feature alignment by optimizing frames to match the gradient of the mesh’s SDF but tends to introduce undesirable singularities along features. In contrast, the frame field generated by NeurFrame effectively guides the quadrilateral mesh to follow feature lines, while avoiding the introduction of excessive singularities or unnaturally placed singularities. At the same time, it maintains globally smooth, low-distortion quadrilateral elements, outperforming QuadWild in overall mesh quality.

The comparison between NeurFrame and other cross-field generation methods is shown in Fig.~\ref{fig:comparison-quad-meshing-cross} and Fig.~\ref{fig:comparison-cross-field}. Benefiting from its continuous representation and volumetric optimization, NeurFrame produces cross fields that align more faithfully with the curvature directions of the surface, even in geometrically complex regions and fine-scale details.
\textcolor{black}{In contrast, competing methods often exhibit noticeable directional inconsistencies, both at global and local levels.}

\subsubsection{Field-guided Hexahedral Mesh Generation}
\thispagestyle{fancy}
\fancyfoot{}

We compare our NeurFrame with existing 3D frame field generation methods~\cite{Huang-2011, Ray-2016} and NeurOcta. To generate the corresponding field-guided hexahedral meshes, we employ the same pipeline and parameters as in our approach. Since CubeCover requires a frame field defined per tetrahedron, whereas the implementations of~\cite{Huang-2011} and~\cite{Ray-2016} define frames per vertex, we adopt the strategy from~\cite{Corman-2019} to compute each tetrahedron's frame $\Othoframe$ as the Karcher mean of the frames $\Othoframe$ at its four vertices $v_i$. Specifically, $\Othoframe$ minimizes $\sum_{i=1}^4 d^2(\Othoframe, \Othoframe_{v_i})$, where $d(\cdot, \cdot)$ denotes the distance on $SO(3)$. For NeurOcta, which outputs a continuously represented neural frame field like NeurFrame, we discretize it by querying the frame at the centroid of each tetrahedron in the volumetric mesh.

\begin{figure}[t]
    \centering
    \includegraphics[width=\columnwidth]{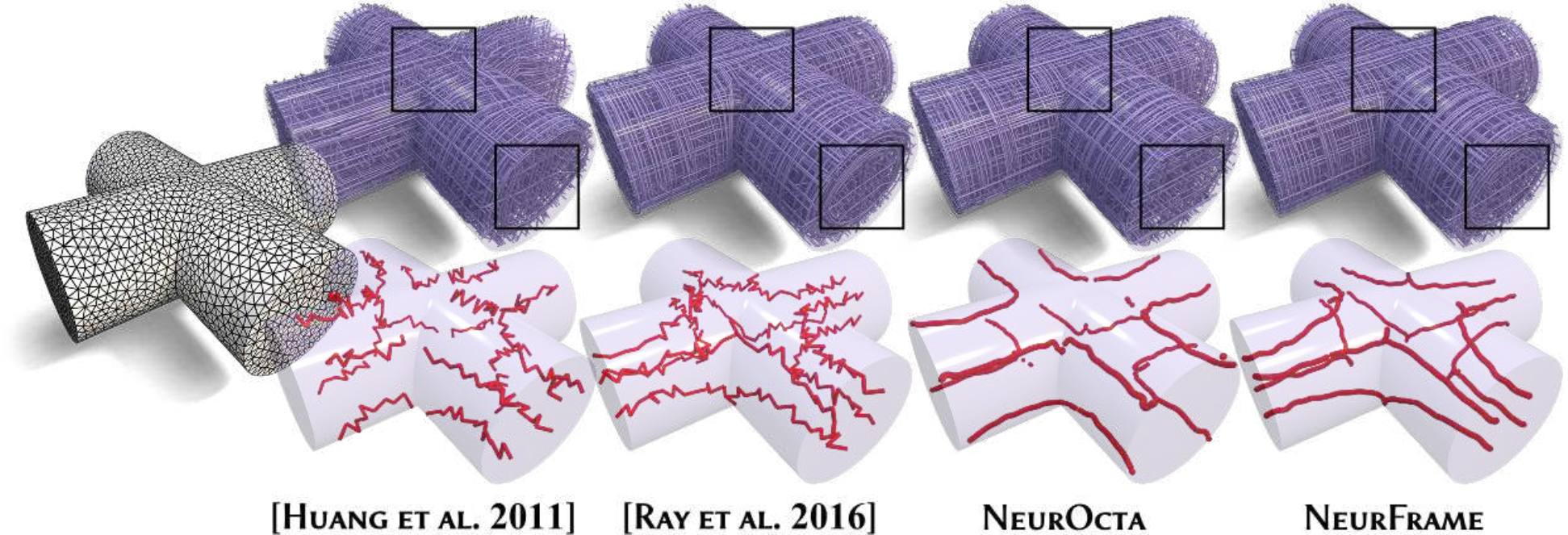}
    \caption{Comparison of frame fields and their singularity graphs. \cite{Huang-2011} fails to balance surface-normal alignment and smoothness. NeurOcta cannot correctly capture cylindrical structures, resulting in a suboptimal singularity graph. \cite{Ray-2016} produces a frame field similar to ours, but NeurFrame’s continuous representation provides superior smoothness.}
    \label{fig:comparison-frame-field}
\end{figure}

\begin{figure}[t]
    \centering
    \includegraphics[width=\columnwidth]{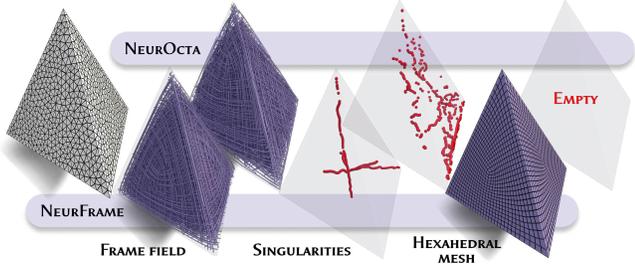}
    \caption{\textcolor{black}{Comparison of boundary alignment on a tetrahedron model. NeurFrame preserves boundary alignment more effectively than NeurOcta, producing a higher-quality singularity structure and enabling a valid hexahedral mesh, whereas NeurOcta fails to generate any valid hexahedral elements in this case.}}
    \label{fig:comparison-neurocta}
    \vspace{-0.2cm} 
\end{figure}

Fig.~\ref{fig:comparison-frame-field} and Fig.~\ref{fig:comparison-hex-meshing} compare frame fields and hexahedral meshes generated by NeurFrame and other methods. NeurFrame produces a frame field that aligns closely with surface normals and varies smoothly across the volume. This improved field quality yields a singularity graph that accurately reflects the structural features of the input model, resulting in a hexahedral mesh that better conforms to the surface and contains higher-quality, lower-distortion elements. Compared to NeurOcta, NeurFrame enforces boundary alignment more effectively, producing better-distributed singularities and enabling successful hexahedral mesh generation; see Fig.~\ref{fig:comparison-neurocta}.

\subsubsection{Runtime Performance}

We compare the per-iteration runtime of neural approaches NeurCross and NeurOcta with that of NeurFrame on our experimental platform. Table~\ref{tab:runtime} reports the average per-iteration runtime (in milliseconds) over 10k training iterations on \textcolor{black}{all models used in this paper}. As shown, NeurFrame achieves the lowest per-iteration runtime while using the same SIREN architecture and number of parameters as NeurOcta, and it is significantly faster than NeurCross, which employs a larger U-Net~\cite{Ronneberger-2015} architecture.

\begin{table}[ht]
\centering
\small
\caption{Average per-iteration runtime (ms) and number of learnable parameters (\#Para.) for neural-based field methods, measured over 10k training iterations on \textcolor{black}{all models used in this paper}.}
\label{tab:runtime}
\begin{tabular}{c|c c c}
    \toprule[1pt]
    Method      & Network       & \#Para.        & Time (ms)\\
    \midrule
    NeurCross   & U-Net         & 4.11M         & 476.60 \\
    NeurOcta    & SIREN         & 266.51K       & 94.78 \\
    NeurFrame   & SIREN         & 266.51K       & 52.87 \\
    \bottomrule[1pt]
\end{tabular}
\vspace{-0.2cm} 
\end{table}

\subsection{Ablation Studies}
\thispagestyle{fancy}
\fancyfoot{}
To evaluate the contribution and sensitivity of each component in our loss function, we conduct ablation experiments by systematically varying $\lambda_{\mathrm{S}}$, $\lambda_{\mathrm{B}}$, and $\lambda_{\mathrm{F}}$ in Eq.~\eqref{loss:total}. First, we assess boundary alignment by reducing $\lambda_{\mathrm{B}}$ from $20$ to $1$ and $0.01$, observing how NeurFrame behaves when the normal constraint is weakened. Next, we scale $\lambda_{\mathrm{S}}$ up (to $10$ and $100$) and down (to $0.1$ and $0.01$) to examine the effect of the smoothness term. Finally, we test the feature alignment term by increasing $\lambda_{\mathrm{F}}$ to $10$ and $100$ to strengthen feature influence. We do not explore further decreases of $\lambda_{\mathrm{F}}$ or much larger $\lambda_{\mathrm{B}}$, as these configurations lead to predictable outcomes: a smaller $\lambda_{\mathrm{F}}$ yields negligible feature guidance, while a larger $\lambda_{\mathrm{B}}$ dominates the smoothness term, effectively reducing the effect of $\lambda_{\mathrm{S}}$. As shown in Fig.~\ref{fig:different-parameters}, our chosen weights achieve a good balance between smoothness, boundary alignment, and feature alignment.

\begin{figure}[t]
    \centering
    \includegraphics[width=\columnwidth]{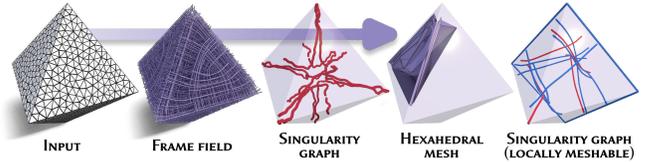}
    \caption{Failure case of our method and the modified singularity graph by~\cite{Liu-2023} on an octahedron model, which remains highly complex and not \emph{globally meshable}, preventing the original algorithm from producing a valid hexahedral mesh.}
    \label{fig:limitation-hexable}
    \vspace{-0.25cm} 
\end{figure}

\section{CONCLUSION AND FUTURE WORK}
\label{sec:conclusion-and-future-work}
In this paper, we presented \emph{NeurFrame}, a neural approach for generating continuous frame fields for structured meshing. NeurFrame can be optimized on a discrete tetrahedral mesh while supporting infinite-resolution queries throughout the domain. Our experiments demonstrate that it produces smooth, boundary-aligned frame fields with fewer and better-distributed singularities, effectively supporting high-quality structured mesh generation. Compared to other self-supervised neural methods for cross-field or frame-field generation, NeurFrame achieves lower computational cost.

However, even though our field is highly smooth and has a simple singularity graph, hexahedral mesh generation can still fail on a simple model; see Fig.~\ref{fig:limitation-hexable}. This occurs because the singularity graph determined by the frame field does not necessarily induce an IGM aligned with it, resulting in a highly degenerate parameterization. Producing a frame field that supports hexahedral meshing (referred to as \emph{globally meshable}) remains an open problem. The state-of-the-art method~\cite{Liu-2023} partially addresses this by modifying the singularity graph to make it \emph{locally meshable}. Nevertheless, even for a simple octahedral model, the modified singularity graph remains highly complex; see Fig.~\ref{fig:limitation-hexable} (rightmost), failing to produce any valid hexahedral meshes. Hence, our future work aims to incorporate the corrected singularity graph into the training process as a target, enabling the neural frame field to be directly meshable.

\bibliographystyle{ACM-Reference-Format}
\bibliography{ref}
\thispagestyle{fancy}
\fancyfoot{}

\clearpage
\thispagestyle{fancy}
\fancyfoot{}

\begin{figure*}[t]
    \centering
    \includegraphics[width=\textwidth]{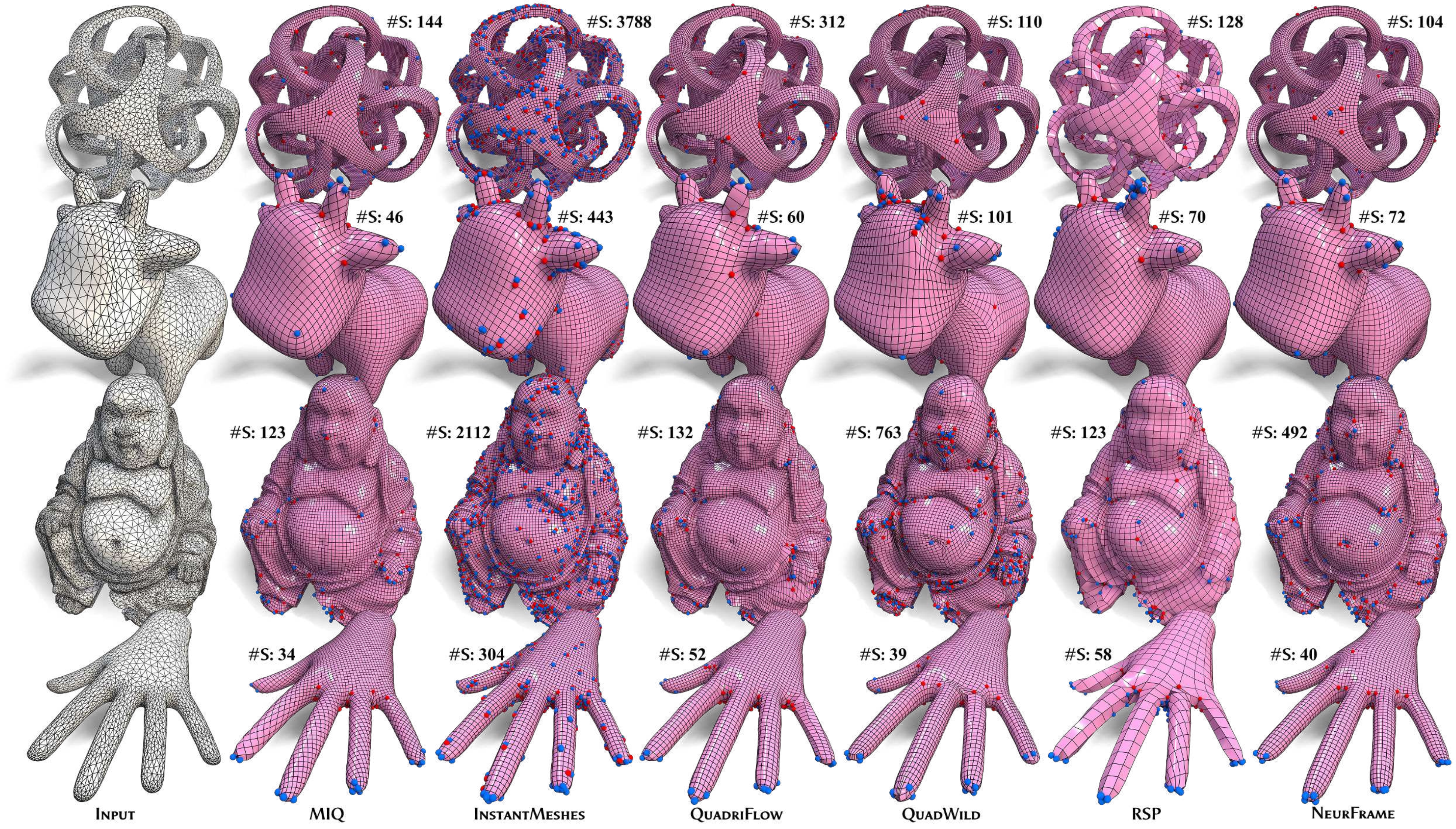}
    \caption{Comparison of NeurFrame with other field-guided quadrilateral mesh generation techniques. Singular points with valence greater or less than four are marked in red and blue, respectively, where \#S denotes the total number of singular points. The RSP method fails to produce meshes with comparable face counts using libQEx, resulting in meshes with reduced face counts.
    }
    \label{fig:comparison-quad-meshing}
\end{figure*}

\begin{figure*}[t]
    \centering
    \includegraphics[width=\textwidth]{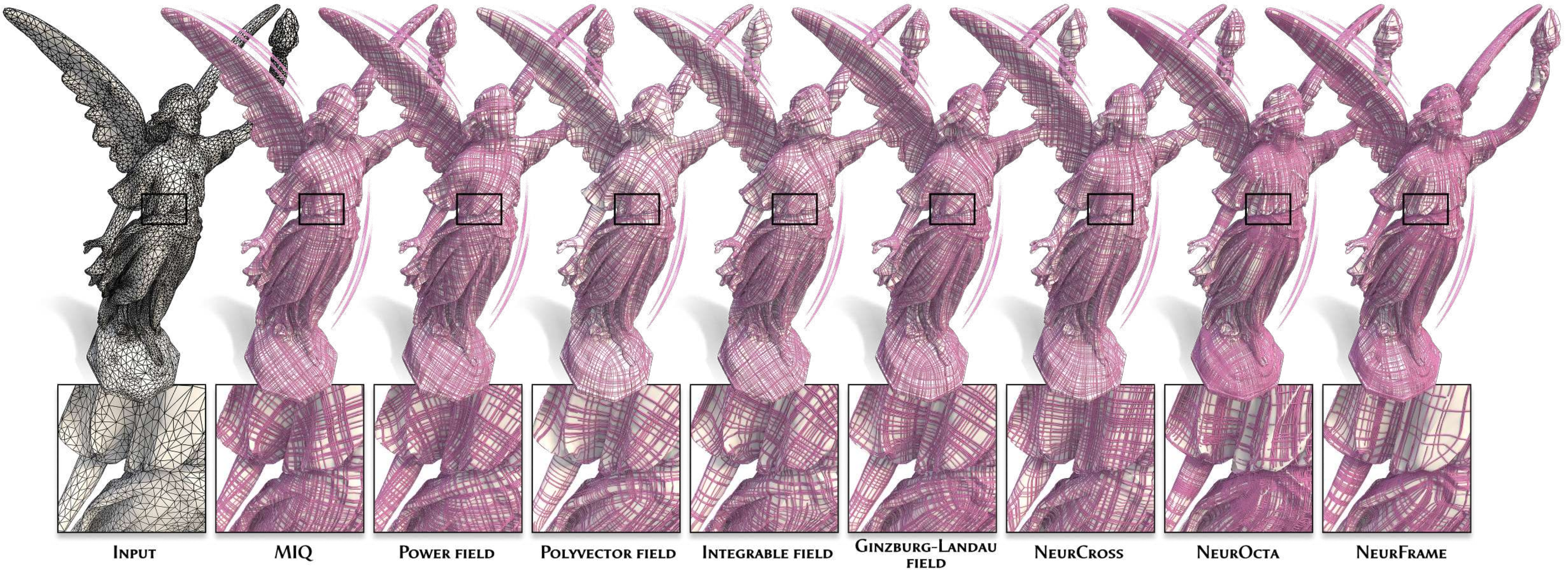}
    \caption{Comparison of cross fields generated by NeurFrame and other methods. On this detailed surface, NeurFrame correctly aligns both global structures (e.g., torso highlighted by the black box and wings) and fine features (e.g., garment hem). In contrast, competitors achieve only partial alignment: MIQ, NeurCross, and NeurOcta align the torso; polyvector, Ginzburg–Landau, and NeurOcta align the wings; power, polyvector, and Ginzburg–Landau align the garment hem.}
    \label{fig:comparison-cross-field}
\end{figure*}

\clearpage
\thispagestyle{fancy}
\fancyfoot{}

\begin{figure*}[t]
    \centering
    \includegraphics[width=\textwidth]{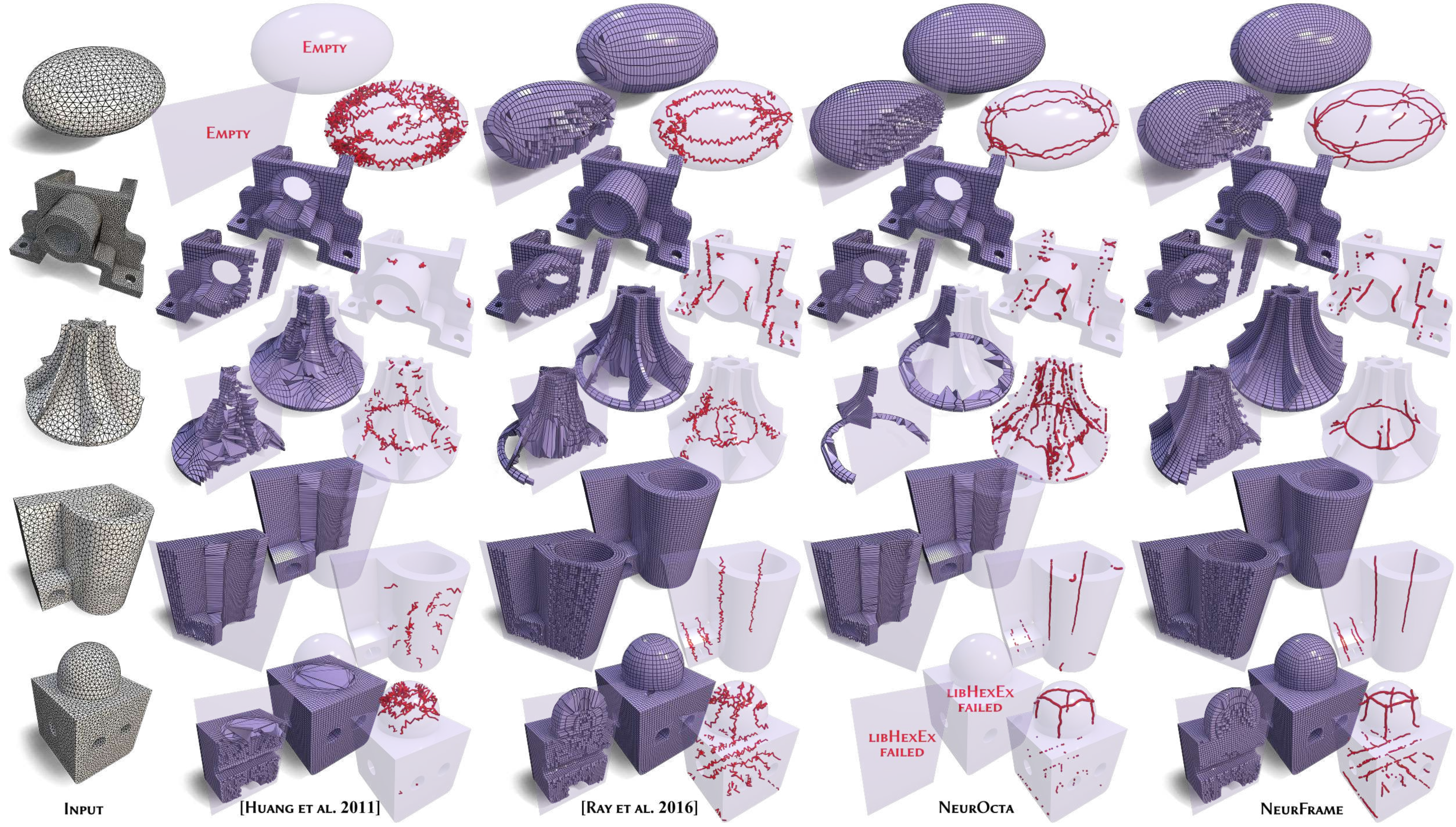}
    \caption{Comparison of hexahedral mesh generation guided by NeurFrame and existing frame field methods. For each model, we show the resulting hexahedral mesh, a cross-sectional view, and the corresponding frame field singularity graph. The frame field produced by \cite{Huang-2011} exhibits disorganized singularity structures, which lead to failed hexahedralization and severely distorted elements. The method of \cite{Ray-2016}, based on a discrete frame field representation, achieves lower smoothness than NeurFrame, resulting in the loss of local geometric structures. NeurOcta shows poor alignment with surface normals, which restricts the output to hexahedral meshes that are topologically close to a cube.}
    \label{fig:comparison-hex-meshing}
\end{figure*}

\begin{figure*}[t]
    \centering
    \includegraphics[width=\textwidth]{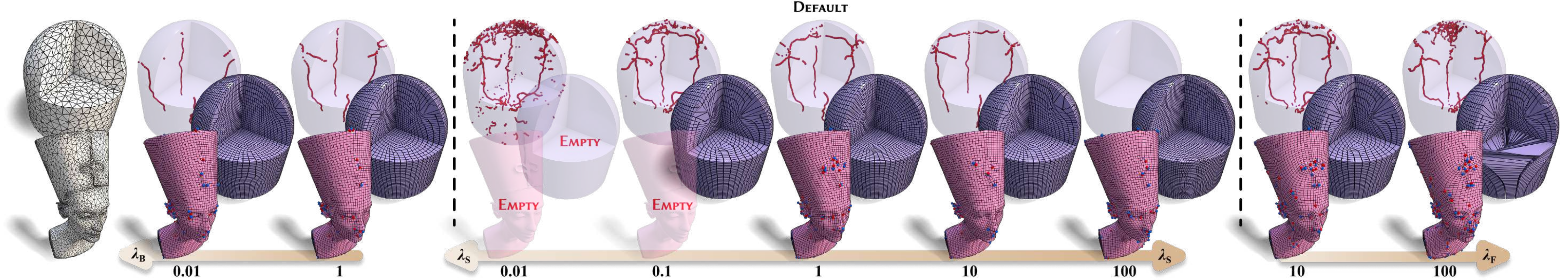}
    \caption{Ablation of loss weights. Varying $\lambda_{\mathrm{S}}$, $\lambda_{\mathrm{B}}$, and $\lambda_{\mathrm{F}}$ shows their effect on smoothness, boundary alignment, and feature alignment. Weak boundary alignment or excessive smoothness leads to fields that deviate from boundaries, reduced smoothness increases interior singularities, and stronger feature alignment can locally distort the field.}
    \label{fig:different-parameters}
\end{figure*}

\clearpage
\thispagestyle{fancy}
\fancyfoot{}

\appendix
\section{IMPLEMENTATION DETAILS}

\subsection{Computation of Singularities in Continuous Fields}

\begin{figure}[t]
    \centering
    \includegraphics[width=\columnwidth]{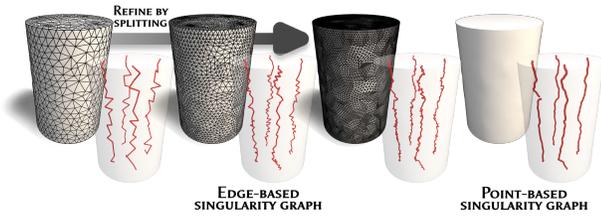}
    \caption{Singularity graph of a frame field. For piecewise linear fields, singularities lie along tetrahedral edges, and mesh refinement improves their approximation. Our continuous frame field is visualized using a point-based representation, capturing singularities directly and smoothly.}
    \label{fig:singularity-computation}
\end{figure}

The singularity graph of a 3D frame field, which forms a network of one-dimensional singular curves throughout the volume, provides an important indicator of field quality and its suitability for subsequent hexahedralization. Conventional methods represent frame fields in a piecewise linear manner, with frames defined at the centroids or vertices of tetrahedra. In this setting, the singularity graph is restricted to tetrahedral edges. For each edge, the dual edges of adjacent tetrahedra form a loop, and the edge is classified as singular if the accumulated frame rotation along the loop deviates from the identity~\cite{Nieser-2011}. Tetrahedral edges typically do not align with the true singularity graph, which makes the extracted edges appear jagged, as shown in the left part of Fig.~\ref{fig:singularity-computation}.
To better visualize the singularity graph of our continuous frame field, we adopt the
\begin{minipage}[b]{\dimexpr\columnwidth-0.3\columnwidth-0.05\columnwidth}
method of~\cite{Solomon-2017} to extract a point-based representation. This approach distributes randomly positioned and oriented equilateral triangles in space and iteratively subdivides those whose boundary rotations deviate from the identity. The cen-
\end{minipage}
\hfill
\begin{minipage}[b]{0.3\columnwidth}
    \centering
    \includegraphics[width=\linewidth]{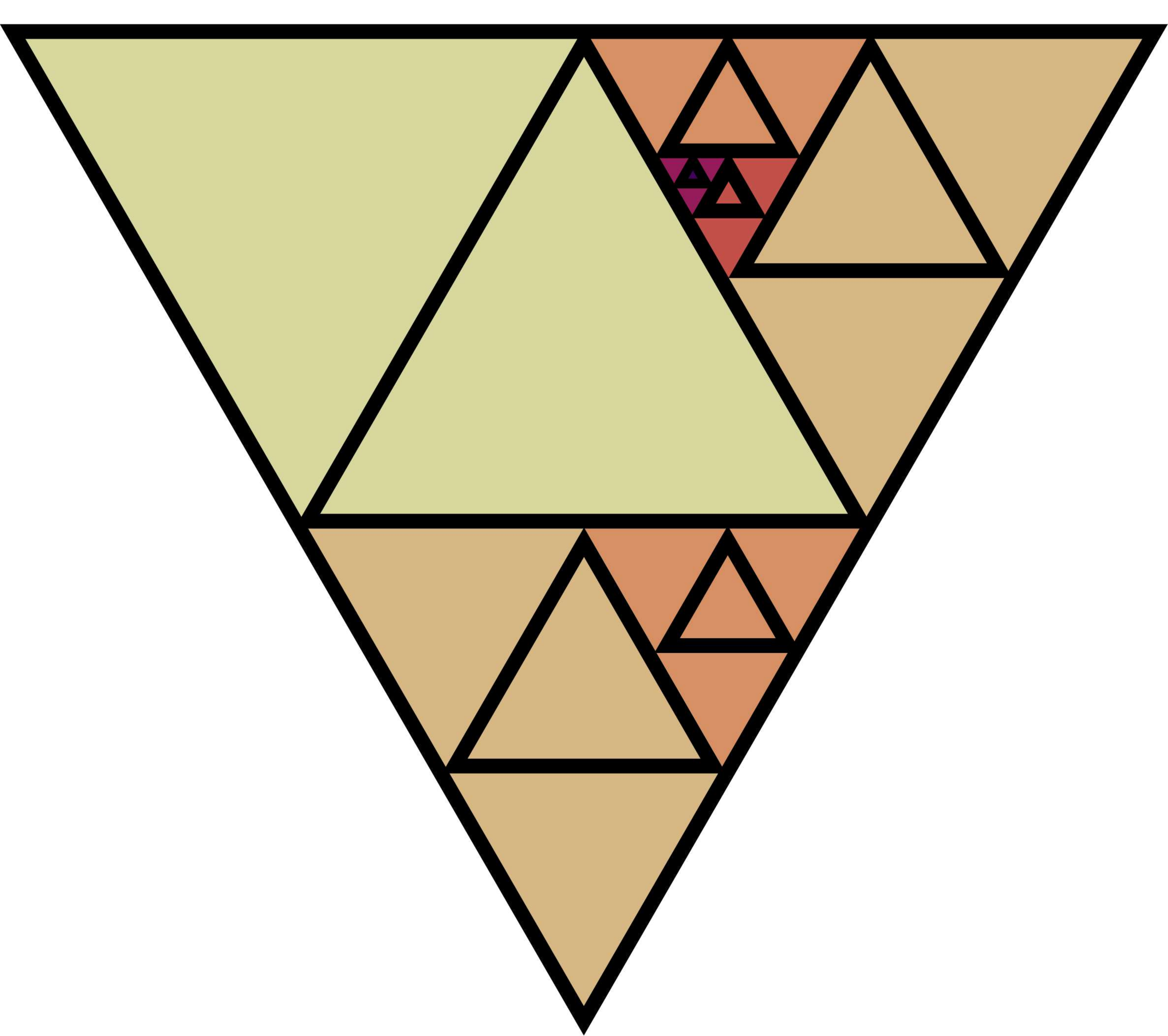}
\end{minipage}
troids of the converged triangles form a dense sampling of the singularity graph, as illustrated in Fig.~\ref{fig:singularity-computation}.

\subsection{Streamline Extraction from Continuous Fields}

Thanks to our continuous frame field representation, the field can be visualized by tracing streamlines from randomly sampled seed points. At each step, a point is advanced along the local frame direction most aligned with its current tracking direction, producing smooth curves that reveal the global field structure (Fig.~1). For surface cross fields, points are projected back onto the mesh after each step, and the surface normal is excluded when selecting the aligned frame direction. This ensures correct propagation along the surface, while occasional right-angle turns naturally reflect the four-fold symmetry of the cross field; see Fig.~1.

\thispagestyle{fancy}
\fancyfoot{}

\end{document}